# Balancing Profit, Risk, and Sustainability for Portfolio Management

Charl Maree*
Center for AI Research
University of Agder
Grimstad, Norway
charl.maree@uia.no

Christian W. Omlin
Center for AI Research
University of Agder
Grimstad, Norway
christian.omlin@uia.no

*Abstract*—Stock portfolio optimization is the process of continuous reallocation of funds to a selection of stocks. This is a particularly well-suited problem for reinforcement learning, as daily rewards are compounding and objective functions may include more than just profit, e.g., risk and sustainability. We developed a novel utility function with the Sharpe ratio representing risk and the environmental, social, and governance score (ESG) representing sustainability. We show that a state-of-the-art policy gradient method – multi-agent deep deterministic policy gradients (MADDPG) – fails to find the optimum policy due to flat policy gradients and we therefore replaced gradient descent with a genetic algorithm for parameter optimization. We show that our system outperforms MADDPG while improving on deep Q-learning approaches by allowing for continuous action spaces. Crucially, by incorporating risk and sustainability criteria in the utility function, we improve on the state-of-the-art in reinforcement learning for portfolio optimization; risk and sustainability are essential in any modern trading strategy and we propose a system that does not merely report these metrics, but that actively optimizes the portfolio to improve on them.

## I. INTRODUCTION

Stock portfolio optimization has been a focal point in financial technology with various solutions proposed including artificial neural networks, support vector machines, random forests, and, more recently, reinforcement learning [1, 2]. The application of reinforcement learning to stock portfolio optimization has generally followed two different approaches: deep Q-learning (DQL) where discretized actions denote buy and sell volumes [3], and policy gradient methods where continuous actions correspond to the distribution of assets in the portfolio [4]. In recent publications, DQL has typically been outperforming policy gradient methods even though discretization is considered disadvantageous [5]. We therefore investigate the cause of the inferior performance of policy gradient methods and propose a solution: replacing gradient descent with a genetic algorithm for parameter optimization. Further, we note that recent studies have typically been using financial returns as the sole performance metric [6]. We propose to include two additional metrics – risk and sustainability – in a novel utility function using the Sharpe ratio and environmental, social, and governance (ESG) score, respectively. While risk is a key element of modern portfolio theory, sustainability is increasingly becoming requisite in financial services. By adding these two metrics to the utility function, we create a system that actively reduces risk while maintaining a sustainable portfolio, thus furthering the state-of-the-art in modern portfolio management.

## II. BACKGROUND AND RELATED WORK

### A. Portfolio Metrics and Market Indicators

The Sharpe ratio is commonly used to quantify the risk-to-reward ratio of a portfolio [7]. It is defined as the expected return in excess of the risk-free return per unit of risk in the portfolio, formally:

$$Sharpe\ ratio = \frac{R_p - R_f}{\sigma_p} \quad (1)$$

Here, $R_p$ and $R_f$ are the expected daily return of the portfolio and the risk-free return respectively, while $\sigma_p$ is the standard deviation of the daily returns of the portfolio. The higher the Sharpe ratio of a portfolio, the better the risk-adjusted performance: a Sharpe ratio less than one is considered sub-optimal by investors, while a ratio greater than one is considered good, greater than two is very good, and greater than three is excellent [8]. The use of the Sharpe ratio in the reward function can significantly increase the return [9].

The environmental, social, and governance (ESG) score is a set of criteria that measure a company's operations for sustainability. It is used by socially aware investors and investment firms to select stocks appropriate to their portfolio, as well as in the finance sector generally; firms such as JPMorgan Chase, Wells Fargo, and Goldman Sachs have all published annual reports that present their ESG performances [10, 11, 12]. While the main purpose of ESG is to provide a measure of sustainable conduct, it may also serve as an indicator of long-term risk; through prioritizing ESG, an investor might be able to avoid companies that conduct high-risk activities with potential future consequences on stock prices. In this study, we use the ESG score reported by Yahoo Finance with a scale of 0-100, where a lower score indicates more sustainable conduct.

Momentum indicators are popular tools used by investors to gauge the strength of a stock. They evaluate the ability of a stock to sustain a rate of price change. Moving average convergence divergence (MACD) is one such indicator which subtracts the 26-day from the 12-day exponential moving average (EMA) – an exponentially weighted moving average, assigning more weight to recent data – of a stock price. MACD is used to predict reversals in trends but is prone to false positives, i.e., it occasionally predicts reversals that do not actually occur. The relative strength index (RSI) is another momentum indicator which is often used in tandem with MACD to mitigate this shortcoming. It uses the magnitude of recent price changes to predict overbought and oversold conditions of a given stock. RSI is calculated as follows:

*Second Affiliation: Chief Technology Office, SpareBank 1 SR-Bank, Stavanger, Norway.

This study was partially funded by The Norwegian Research Council; project nr 311465.

$$RSI = 100 - \frac{100}{1 + \frac{P_x}{N_x}} \quad (2)$$

Here, $P_x$ and $N_x$ are the averages of the positive and negative close prices respectively, for a period of $x$ days. The RSI value lies between 0 and 100, and the typical interpretation is that values below 30 and above 70 indicate the stock being oversold and overbought, respectively. Studies have shown that MACD and RSI can increase returns for stock trading. [13, 14].

The final indicator we used is drawdown, specifically daily drawdown (DDD) and maximum drawdown (MDD). While the former is calculated as the scaled difference between the current ($P_{current}$) and maximum ($P_{max}$) stock prices for a given period, the latter is the scaled difference between the minimum ($P_{min}$) and maximum ($P_{max}$):

$$DDD = \frac{P_{current} - P_{max}}{P_{max}} \quad (3)$$

$$MDD = \frac{P_{min} - P_{max}}{P_{max}} \quad (4)$$

Drawdown is one of the most widely used indictors of risk and is a measure of *downside* volatility, in contrast to the Sharpe ratio which is a measure of volatility in general [15]. It is therefore especially useful to, e.g., short-term investors to whom *upside* volatility is not of paramount concern.

### B. Non-Stationarity in Reinforcement Learning

In reinforcement learning, agents learn policies by maximizing expected cumulative rewards [16]; the value of each state in a Markov decision process (MDP) is the discounted sum of rewards of future states, formalized by the Bellman equation [17]:

$$V(s) = \max_{a \in A(s)} \sum_{s'} P(s' \mid s, a)\bigl(R(s, a, s') + \gamma V(s')\bigr) \quad (5)$$

Here, $V(s)$ is the value of state $s$, $P(s' \mid s, a)$ is the probability of transitioning to state $s'$ given state $s$ and action $a$, $R(s, a, s')$ is the reward for action $a$ in state $s$ transitioning to sate $s'$, and $\gamma \in [0,1]$ is the discount rate which reduces the weight of future rewards. The value of a state is the maximum discounted reward for all possible actions for that state, $A(s)$. While Equation (5) is the general Bellman equation for stochastic MDPs, deterministic MDPs will have the transition probability distribution $P(s' \mid s, a)$ reduced to one. Furthermore, stochastic systems may either be stationary or non-stationary. Unlike stationary systems which have constant transition probability distributions, non-stationary systems have proven problematic for traditional reinforcement learning methods [18]. A relevant example of a non-stationary MDP is a multi-agent system where multiple independent agents act on the same environment resulting in unstable state transition probabilities caused by the changing policies of the other agents during training [18].

### C. MADDPG for Stabilizing a Multi-Agent System

Multi-agent deep deterministic policy gradient (MADDPG) was introduced to address the inherent non-stationarity of multi-agent systems [18]. In their paper, the authors demonstrated an increasing variance in policy gradients with an increasing number of agents. They extended deep deterministic policy gradient (DDPG) in which the parameters $\theta$ of the optimum policy $\pi^\star$ are determined through maximizing the objective function $J(\theta) = E_{s \sim p^\pi, a \sim \pi(\theta)}[R]$, where $p^\pi$ is the state distribution and $\pi(\theta)$ is the policy according to parameters $\theta$. They formalized the gradient of the objective function for deterministic policies ($\mu_\theta : S \mapsto A$) as:

$$\nabla_\theta J(\theta) = E_s\bigl[\nabla_\theta \mu_\theta(a \mid s) \nabla_a Q^\mu(s, a)\bigr|_{a = \mu_\theta(s)}\bigr] \quad (6)$$

In DDPG, $\mu_\theta(a \mid s)$ is modelled by an actor network which predicts the best action given a state, while the reward function $Q^\mu(s, a)$ is modelled by a critic network which estimates the value of a state-action pair. These networks experience high variance in their policy gradients when used in multi-agent settings, as the actions of other agents are absent in the loss function while the rewards depend on these actions [18]. The authors in [18] mitigated this problem by extending the critic $Q^\mu(s, a)$ to consider the actions of all agents: $Q^\mu(s, a_i, i \in \{1 \dots N\})$, where $N$ is the number of agents.

### D. Genetic Algorithms for Parameter Optimization

In general, genetic algorithms (GA) solve problems by evolving a population of individuals $a_i, i \in \{1 \dots N\}$, each with a set of parameters $\theta_i$. At each generation $g$, a fitness score $F(a \mid \theta_{i,g})$ is calculated for each individual through measuring their performance at solving a given problem. Typically, the best performing individual is carried over to the next generation ($g + 1$), while the top $k < N$ individuals are used to generate a new batch of $N - 1$ individuals, such that the size of the population remains constant. This new population is generated either through parameter mutation – where parameters are altered through crossover-mutation between parents' parameters – or through the addition of Gaussian noise: $\theta_{g+1} = \theta_g + \sigma \epsilon$ where $\epsilon \sim \mathcal{N}(0, I)$ and $\sigma$ is a hyperparameter which roughly corresponds to a learning rate. In [19], the authors used the addition of Gaussian noise to evolve the parameters of a neural network and found that it outperformed both DQL and gradient-based methods at playing games[1]. In another study, the authors used GA to evolve the parameters of a single agent system and showed that it outperformed DDPG in moving a physical robotic arm [20].

### E. Reinforcement Learning for Stock Portfolio Optimization

In stock portfolio optimization, a trader continuously redistributes funds between a selection of stocks. Risk-aware traders structure their portfolios to optimize risk for a given expected return [21]; one approach is portfolio optimization using reinforcement learning. In [22], the authors compared the performance of different single-agent policy gradient methods on an MDP structured as follows:

- *State:* the close-price history, high-price history and a wavelet transform of the close-price for each of six stocks for a given time window.

---

[1] Games are a popular application for reinforcement learning as they facilitate learning on high-dimensional input data akin to human sensory input such as vision [29].

- *Action:* a continuous daily distribution of funds across the six stocks.
- *Reward:* $\log(\Delta P) + S$, where $\Delta P$ is the daily change in portfolio value, and $S$ is the Sharpe value.

It could be argued that this approach does not appropriately weight risk for all types of investors; certain investors might be more risk-averse than others, e.g., individuals in different stages of their lives. The authors stated that even though DDPG was their best performing method, it performed rather poorly and frequently ended in local minima. Their best performing scenario with a careful stock selection achieved approximately 25% annual returns.

Similarly, the authors in [23] presented a DDPG-based method for trading a selection of 8 stocks. They used LSTM networks for the critics and feed-forward networks for the actors. Their state consisted of daily stock prices, RSI, stock positions, and the portfolio value. Their rewards were simple daily returns, and their actions were the continuous distribution of stock positions in the portfolio. They reported compound annual return of 14% and a Sharpe ratio of 0.6 over a period of 11 years.

In [24], the authors presented a multi-agent DQL system that traded four different crypto currencies – Bitcoin (BTC), Litecoin (LTC), Etherium (ETH), and Ripple (XRP). In this system, each agent traded a single asset and the MDP was formalized as follows:

- *State*: the close price for each asset at the given time step.
- *Action*: $2 \times 30$ discretized bins for buy and sell respectively, and one action to hold, totaling 61 actions.
- *Reward*: two reward functions were tested: a simple sum of financial returns and a weighted sum of the returns and the Sharpe ratio.

The weighting between the returns and Sharpe ratio was a hyperparameter – an improvement over [22] as this potentially allows for different strategies depending on the investor's appetite for risk. The authors reported that the second reward function yielded better results. They reported daily returns between 2.0% and 4.7%, while the best annualized Sharpe ratio achieved was 3.2. This system clearly performed better than the ones in [22] and [23], which could be related to the nature of the optimizer in a discretized action space; DQL does not rely on policy gradients and is therefore not susceptible to local minima. Another difference is that this study used multiple agents, i.e., one agent per stock. It could be argued, however, that these agents were simply clones that fulfilled the same role given the same observations and rewards, and that they could learn neither unique behaviors nor cooperation.

The DQL system presented in [25] divided the portfolio optimization problem into timing and pricing elements which resulted in two types of agents: signal agents and order agents, respectively. Additionally, each of these two types of agents were concerned with either buying or selling of assets, which resulted in four individual agents. Agents had individual *state* observations: while buy and sell signal agents received a history of asset prices, the sell agent also received information about potential profit using the next-day stock price. Further, the buy and sell order agents' observations were market indicators – the Granville indicator [2] and Japanese Candlesticks[3]. The *action* spaces for the four agents consisted of buy and sell signals sent from the signal agents to the appropriate order agents which in turn generated discrete buy or sell volumes. The *reward* function $R \in [0,1]$ was the normalized difference between the selling or buying price and the high or low price of the next day, respectively. Though the authors presented their results in percentage profit over a 4.5-year test period (1138.7%), we calculated their compound annual return for their best-case scenario as 74.9%. They did not report a Sharpe ratio for their optimized portfolio.

In summary, discretized DQL systems typically outperform policy gradient systems for stock portfolio optimization. We hypothesized that this could be attributed to the nature of the policy gradients; flat policy gradients and local minima pose challenges for gradient-based optimizers [18]. We therefore replaced gradient descent with a genetic algorithm for parameter optimization to eliminate gradient-based optimization problems while maintaining a continuous action space. A continuous action space is desirable because stock trading is not inherently discrete, and discretization adds an unnecessary level of abstraction [5]. Finally, there has not – to the best of our knowledge – been any published reinforcement learning portfolio optimization system that used ESG in its utility function. This is a significant oversight since sustainable investing is pivotal to a more sustainable society [10]. We therefore address this by incorporating ESG in our utility function.

III. EMPYRICAL METHODOLOGY

A. Data

We used market data as reported by Yahoo Finance for a selection three of stocks from the DOW30 index: The Goldman Sachs Group, Inc. (GS), The Procter & Gamble Company (PG), and 3M Company (MMM). For our training and testing periods, these three stocks had constant ESG risk scores of 28.12, 25.10, and 34.88, respectively. However, it is possible that ESG sores can change in time according to changes in companies' operations and our system is designed to cope with such changes. We used the asset close prices for a period of two years in training (shown in Fig. 1a) and the following year in testing (shown in Fig. 1b); it is injudicious in stock portfolio optimization to not have a separate test set, firstly because trading will always happen on unseen data, and secondly because typical MDPs for stock portfolio optimization are non-stationary and therefore render reinforcement learning agents susceptible to overfitting [26, 27].

B. Design of Markov Decision Process

Many studies have been avoiding policy gradient methods for portfolio optimization by discretizing action spaces. However, the portfolio optimization problem is not inherently discrete and continuous action spaces are therefore considered preferable [5]. In this study, we used a triple agent system and the following MDP with a continuous action space:

---

[2] The Granville indicator is a set of eight conditions of a stock price in relation to its moving average, e.g., a bullish breakthrough is when the stock price crosses the moving average in an upward trend. It indicates buying or selling conditions.

[3] Japanese candlesticks consider four daily price points: open, close, high, and low. A stock is considered either bearish or bullish depending on the difference between open and close prices while the high and low prices indicate daily volatility.

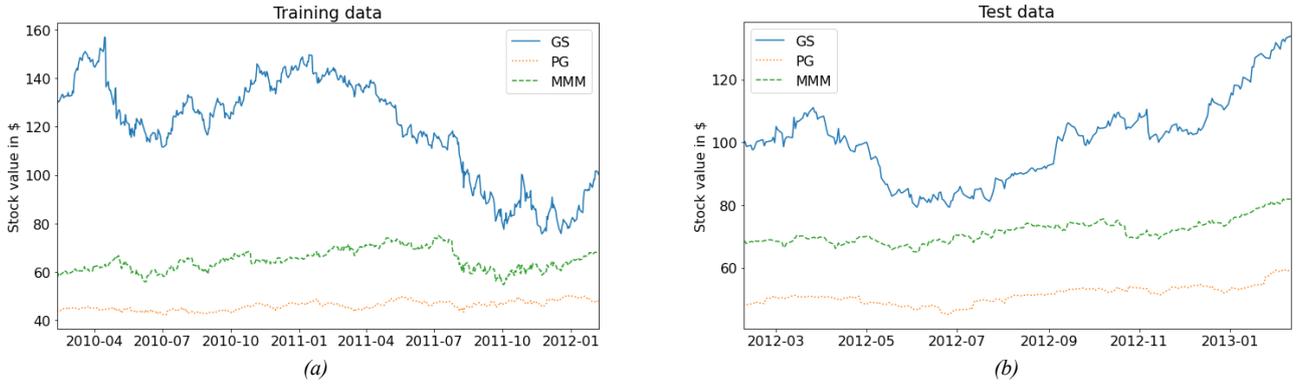

*Fig. 1 The stock price data used for (a) training and (b) testing purposes. Market data is shown for The Goldman Sachs Group, Inc. (GS), The Procter & Gamble Company (PG), and 3M Company (MMM). Market conditons were slightly different between these two datasets, e.g., GS experienced an overall decrease in stock price for the training period, but an overall increase during the testing period.*

- Our *state* was represented by 18 values: for each of the three stocks a normalized stock price (with subtracted mean, scaled to unit variance), MACD, RSI, DDD, MDD, and the difference between the 20-day and 5-day EMA's.
- The *action*-spaces of our first two agents (profit agent and risk-averse agent) were the continuous distributions of positions for the three stocks and one for holding cash, i.e., there were four values per action ($A_1$ and $A_2$ respectively where $|A_i| = 4$, $A_{i,j} \in [0,1]$, $\sum_{j=1}^{4} A_{i,j} = 1, i \in \{1,2\}$ ). The third action (the sustainability action) consistently selected best performing stock with respect to ESG, e.g., $A_3 = [0,1,0,0]$ while PG had the lowest ESG score. The final agent's (manager agent) action was the weighting between the three actions (profit, risk, and sustainability): $A_t = \sum_{i=1}^{3} \beta_i A_{i,t}$, where $A_t$ was the total action sent to the environment at time-step $t$ and $\beta_i \in [0,1], i \in [1,3]$ was the third agent's action.
- The rewards were unique to each agent: the profit agent's reward was the change in portfolio value from time-step $t$ to $t+1$: $r_{1,t} = \Delta P|_t^{t+1}$; it was only concerned with profit. The risk-averse agent's reward was the Sharpe ratio for a moving window of 20 days: $r_{2,t} = Sharpe(t - 20 \rightarrow t)$ if $t \geq 20$, else $0$; it was concerned with risk and the variability of daily returns. The manager agent's reward was a linearly weighted function of the rewards of the first two agents and the mean ESG score for the portfolio: $r_4 = \sum_{i=1}^{3} \omega_i r_{i,t}$, $\sum_{i=1}^{3} \omega_i = 1$, $\omega_i \in [0,1]$ where $r_{3,t} = -\sum_{i=1}^{3}(x_i \cdot ESG_i)$ where $x_i$ and $ESG_i$ are the position and ESG score of stock $i$ and the weighting parameters $\omega_i, i \in [1,3]$ were hyperparameters which we tuned to the values of 0.7, 0.2 and 0.1, respectively; the manager agent weighed the recommended actions from the other agents to achieve balanced rewards given a tunable prioritization between risk, reward and sustainability.

Finally, we calculated our market indicators as follows: We used standard periods of 14 days in Equation (2) for RSI and 26 days in Equations (3) and (4) for DDD and MDD. We annualized the Sharpe ratio by assuming 252 trading days per year $Sharpe_{annual} = \sqrt{252} \cdot Sharpe_{daily}$ where $Sharpe_{daily}$ was calculated from Equation (1) with a risk-free return equal to zero; we assumed risk-free returns were negligible which is not an unusual assumption with consistently low interest rates for our selected time period.

### C. Design of Agents

We compared two systems of three agents acting on the MDP described above: a MADDPG system (as described in [18]), and a system using a genetic algorithm to optimize the parameters of the deep neural networks of the agents. The MADDPG agents each had two feed-forward neural networks, one for the actor and one for the critic. The actor networks' inputs were complete observations of the state described above, while their outputs were the actions as described above. The critic networks' inputs were a complete observation of the state plus the actions of *all* agents, while their outputs were the estimated value of the current state. The hidden layers were the same for all networks: two fully connected layers of 64 nodes each, followed by a softmax activation for the actor networks and no activation for the critic networks. We tuned the learning rate to 0.001, discount factor ($\gamma$) to 0.99, and target-network update parameter ($\tau$) to 0.01 for all agents. The training batches were relatively large (256 samples) to mitigate the effects of the observed flat policy gradients. Each training run consisted of 5,000 iterations, each with one data collection episode and three training batches, and the replay buffer was sized to store the transition trajectories for two episodes. The system based on genetic algorithms used identical actor networks to that of the MADDPG system, without the need for critic networks. For a tuned population size of 200, we mutated the fittest 10% of each generation using random mutation and a gaussian noise multiplier $\sigma$ tuned to 0.3 while carrying over the fittest individual unmutated. For both systems, hyperparameter tuning was done through a standard one-at-a-time parameter sweep.

### IV. RESULTS

In Table 1, we show the results of our experiments compared to that of published work on both continuous and discretized action spaces for stock portfolio optimization.

TABLE 1 RESULTS FROM OUR GENETIC ALGORITHM (GA) AND MADDPG SYSTEMS COMPARED TO TYPICAL DDPG AND DQL SYSTEMS.

| System | Returns* | Sharpe ratio* | ESG* |
|---|---|---|---|
| Our GA | 70.4% ± 6.8% | 3.15 ± 0.22 | 26.9 ± 1.2 |
| Our MADDPG | 27.9% ± 9.5% | 1.28 ± 0.42 | 29.6 ± 0.3 |
| DDPG | 25% [22] | 0.6 [23] | - |
| DQL | 74.9% [25] | 3.2 [24] | - |

*Ranges are for 95% confidence intervals.

Our MADDPG returns were in line with the returns reported in the single agent DDPG system in [22], while the returns of the better performing DQL system in [25] were within the 95% confidence interval of our GA system, making them essentially the same. Further, our GA system outperformed our own MADDPG system in terms of sustainability, with a superior ESG score of 26.9 compared to 29.6. Our GA system also achieved low risk, with a Sharpe ratio of 3.15 which is typically considered "excellent", while a Sharpe ratio of 1.28 as achieved by the MADDPG system is considered merely "good" [8]. Finally, while [24] reported a similar Sharpe ratio to our system, it was merely a reported metric whereas our system took an active approach to minimizing risk. Our system could thus match [24] in terms of risk *and* [25] in terms of profit, while each of these systems were inferior to ours otherwise. Therefore, though we did not strictly outperform DQL systems in terms of pure financial returns, the fact that we can match their financial returns while offering reduced risk and a sustainable portfolio leads us to claim that our solution is an improvement over the state-of-the-art.

In Fig. 2 we show two typical portfolios held by the two systems during testing. While the GA system quickly achieved a positive portfolio value, the MADDPG system fluctuated around the break-even line for the first half of the episode. Only when the market entered a bullish state after roughly 180 days did we observe a markable increase in the MADDPG portfolio value. This increase was observed much earlier for the GA system – after about 100 days. The fact that the GA system held positions in the GS stock during evaluation, despite this stock having had a mostly downward trend in the training data, suggests that it had learned to interpret market signals as opposed to simply holding the stock that performed best during training. The GA system also responded better to market fluctuations by, for example, taking a position in PG while GS showed bearish signals between ca. 40 and 60 days and choosing to hold cash at times when the MADDPG system did not. The two systems had clearly learned different strategies, reiterating that for at least one of them the optimum policy remained elusive.

We verified that the substantial difference in performance between the MADDPG and GA systems was due to the nature of the MADDPG system's policy gradients. In Equation (6) we showed that the objective function of a MADDPG system is expressed in terms of the parameters of the actor ($\mu$) and critic ($Q$) networks. Fig. 3 illustrates the steepest negative gradients of each of the actor and critic networks during training and gives an indication of how well an optimizer may perform at gradient descent; if all gradients are flat – or close to zero – then the optimizer has no indication of how to adjust the weights. From this figure, we observed that while the three *critics* appeared to have had sufficient gradients to perform gradient descent, the *actors* all experienced flat gradients. This suggests that the critics were able to learn the values of states, but the agents were not able to effectively use these values to find optimum policies. This might be due to the critics having had a more holistic view of the state action space, as intended by the authors of MADDPG [18]. We therefore conclude that the optimum policy had remained elusive to the MADDPG system, as substantiated by the higher returns achieved using the GA system.

Finally, in Fig. 4 and Fig. 5, we show the actions taken by the *individual agents* of the two systems. The agents of the GA system clearly took more distinct roles, with the risk-averse agent frequently voting to hold cash and generally avoiding the most volatile of the three stocks (GS), which the profit agent mostly favored. Interestingly, this clear separation of responsibility was not evident in the behavior of the MADDPG agents which acted more haphazardly. In future work, we intend to more closely inspect the behaviors of the agents and we aim to characterize them and extract explanations for their actions.

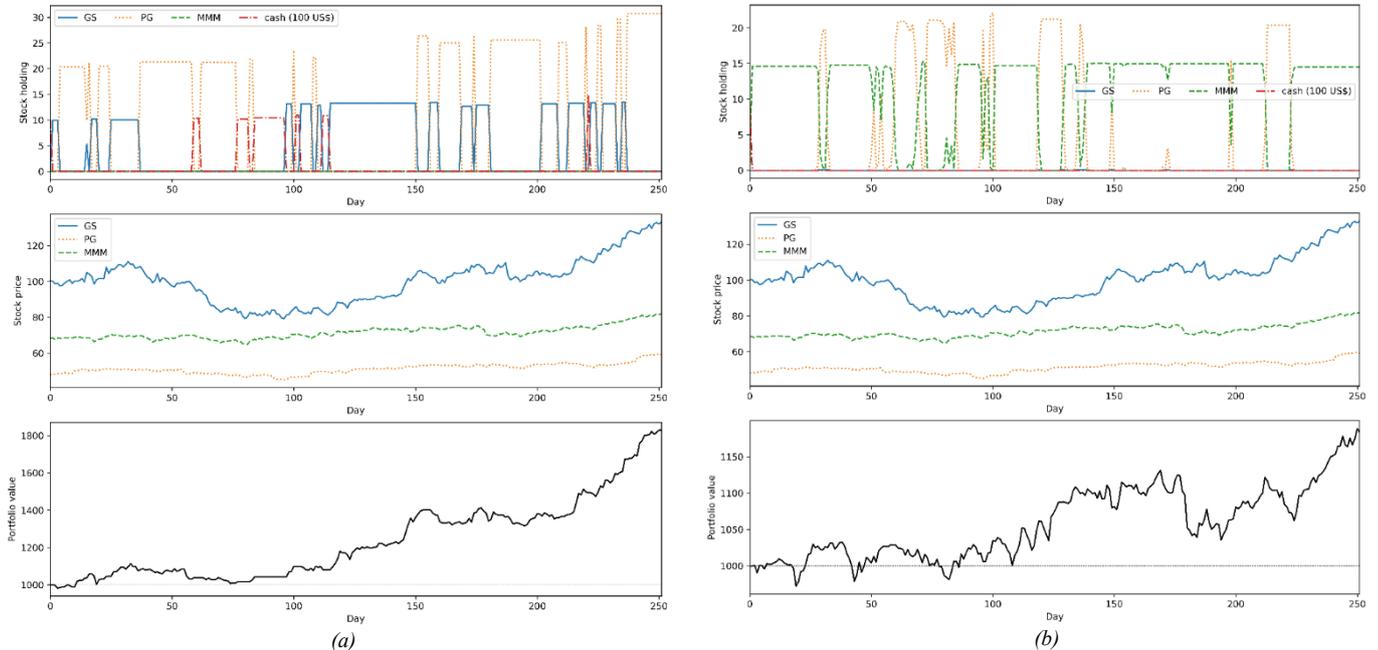

*Fig. 2 Two typical portfolios when trading stocks on test data using (a) the genetic algorithm system and (b) the MADDPG system of traders. The upper row shows the daily positions of assets – GA, PG, MMM, and cash – for each system, the middle row shows the daily close prices, and the bottom row shows the total portfolio values of each system. While the system in (a) hardly has negative portfolio values and finally achieves a return of 83% (Sharpe ratio = 3.4 and ESG = 26.0), the system in (b) frequently has negative portfolio values in early stages and finally yields a return of 19% (Sharpe ratio = 1.2 and ESG = 27.8.)*

## V. CONCLUSIONS AND DIRECTIONS FOR FUTURE WORK

In this study, we defined the problem of stock portfolio optimization in terms of reinforcement learning and designed a multi-agent system with a continuous action space. From published works we showed that, for stock portfolio optimization, DQL has typically been outperforming policy gradient methods despite being limited to discretized actions. We then showed that, for our problem, this was due to flat policy gradients inhibiting gradient descent from finding the optimum policy. Since continuous action spaces are nevertheless considered preferable to discretization in stock portfolio optimization, we overcame the policy gradient problem by replacing gradient descent with a genetic algorithm for parameter optimization. We showed that this method outperformed MADDPG in a three-agent system trading a selection of three stocks from the DOW30 index. Furthermore, our agents were rewarded not only for financial returns, but also for risk (via the Sharpe ratio), and sustainability (via ESG). While risk is key in modern portfolio theory, sustainability is increasingly becoming requisite in modern trading strategies. It is therefore pivotal that state-of-the-art solutions not only support reporting of such metrics, but that they actively optimize portfolios to improve on them. Our main contribution is therefore the inclusion of the Sharpe ratio and ESG in the utility function for portfolio optimization, while matching state-of-the-art solutions in terms of financial returns. We claim that is not necessary to outperform current solutions in terms of financial returns since our solution offers the same returns with reduced risk in a sustainable portfolio, making it an improvement on the state-of the-art.

Our ultimate objective is the development of personalized digital financial advisors using customer micro-segmentation [28]. These financial advisors will recommend, in an explainable way, an optimum allocation of funds given a personal budget and a portfolio of financial products and services. We therefore intend to address the explainability of our system in future work by characterizing, explaining, and predicting an agent's policy based on the history of past trades.

## ACKNOWLEDGMENT

We acknowledge Phillip Tabor for his implementation of the MADDPG algorithm. Some of the code for this paper was adapted from his library.

# VI. APPENDIX

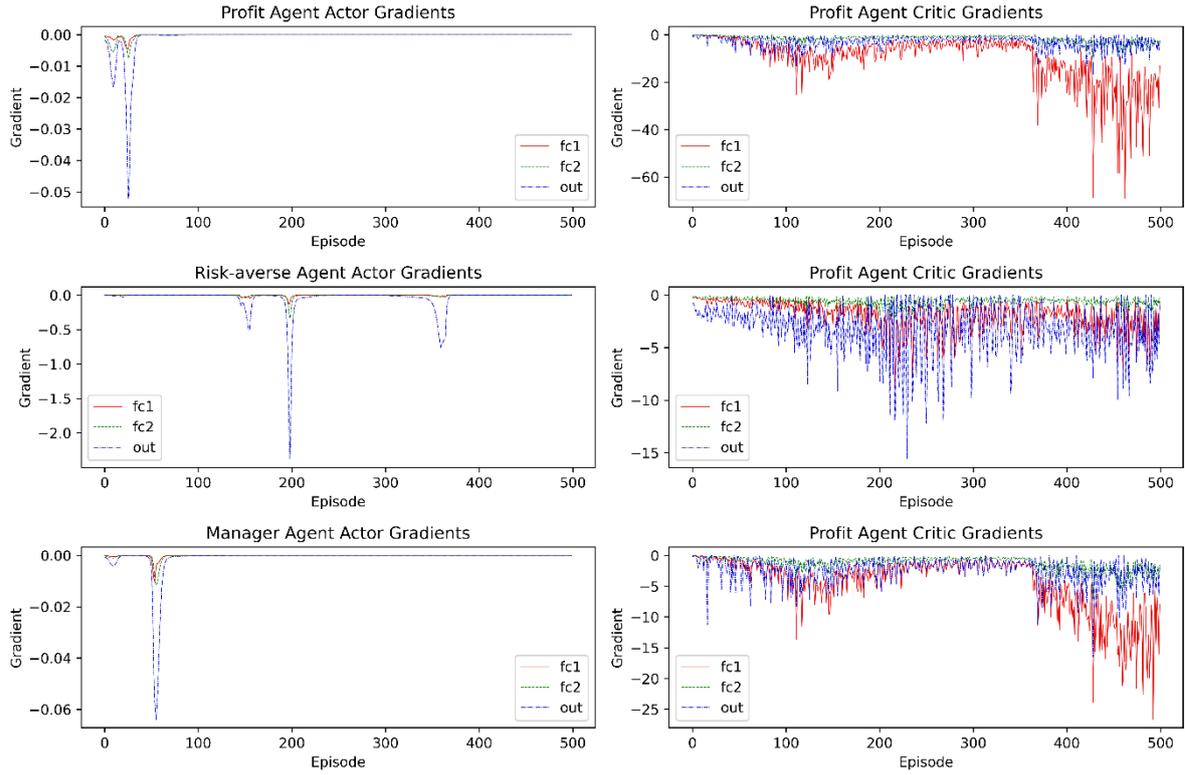

*Fig. 3 The steepest policy gradients of the three MADDPG agents' actor and critic networks for the first 500 training episodes. Each datapoint shows the largest negative component of the gradients of the weights for each of the fully connected (fc1, fc2) and output layers (out). While the critic networks have workable gradients, the gradients for the actor networks are mostly flat throughout training.*

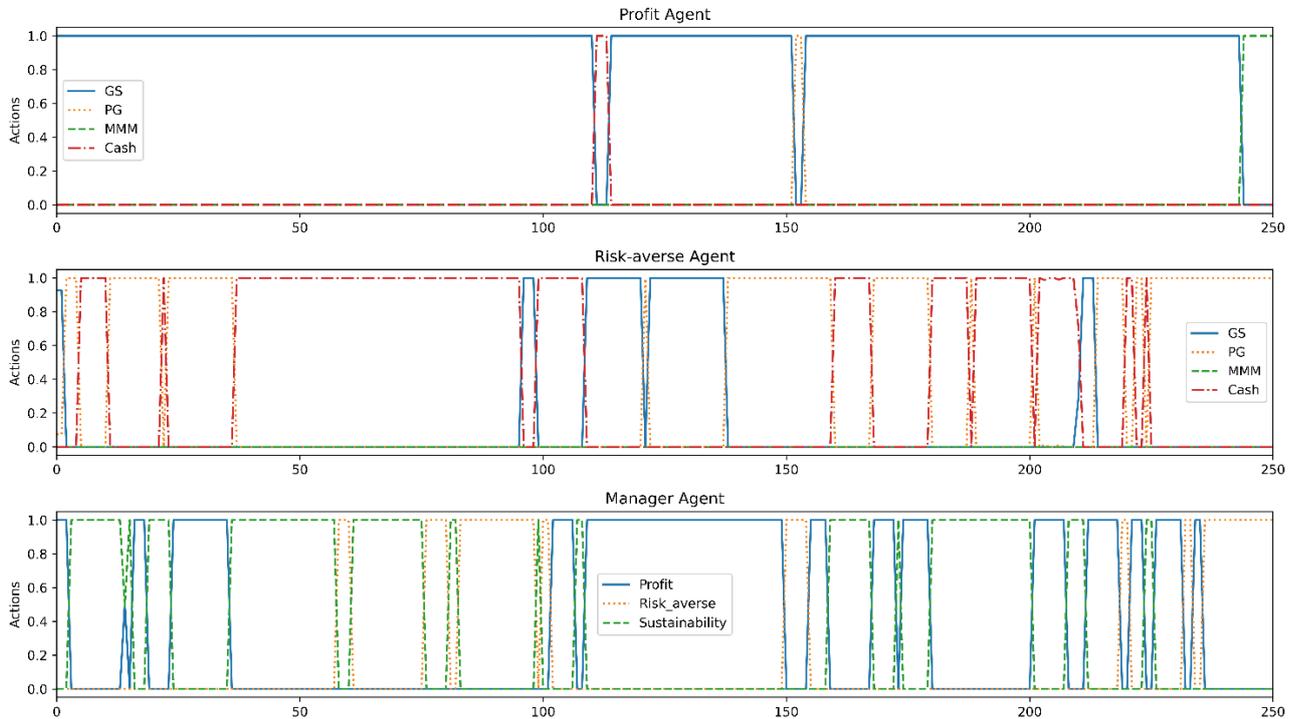

*Fig. 4 The agents' actions in a typical GA system during testing. The first plot shows the profit agent frequently voting to hold GS, while the second plot shows the risk-averse agent more most frequently voting to holding cash. The manager agent's role was to choose the weighting between the other agents' votes; the last plot shows it frequently varying between all three objectives: profit, risk, and sustainability.*

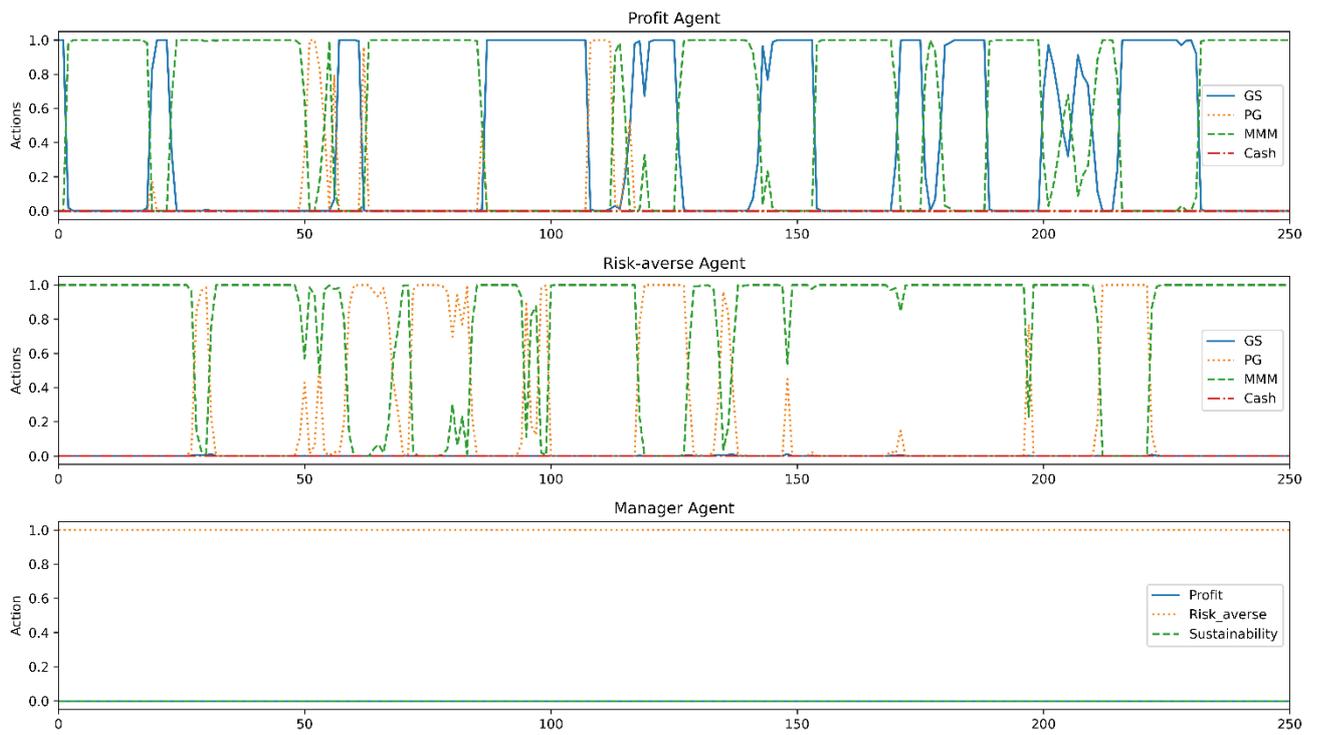

*Fig. 5 The agents' actions in a typical MADDPG system during testing. The fist plot shows the profit agent's vote varying mostly between GS and MMM, while the second plot shows the risk-averse agent voting mostly for MMM, interestingly without ever holding cash. The manager agent's role was to choose the weighting between the other agents' votes; the last plot shows that it always chose to accept the vote of the risk-averse agent.*